%% file: 5624.tex
\documentclass{aa}
\usepackage{natbib}
\usepackage{txfonts}
\usepackage{graphicx}

\bibpunct{(}{)}{;}{a}{}{,} 

\DeclareSymbolFont{UPM}{U}{eur}{m}{n}
\SetSymbolFont{UPM}{bold}{U}{eur}{b}{n}
\DeclareMathSymbol{\umu}{0}{UPM}{"16}

\begin{document}

\title{Molecular dark matter in galaxies}

\author{T.~A.~Bell\inst{1}
\and T.~W.~Hartquist\inst{2}
\and S.~Viti\inst{1}
\and D.~A.~Williams\inst{1}}

\institute{Department of Physics and Astronomy, University College London, Gower Street, London WC1E 6BT, UK\\
\email{tab@star.ucl.ac.uk}
\and School of Physics and Astronomy, University of Leeds, Woodhouse Lane, Leeds LS2 9JT, UK}

\date{Received 17 May 2006 / Accepted 17 July 2006}

\abstract
{Clouds containing molecular dark matter in quantities relevant for star formation may exist in minihaloes of the type of cold dark matter included in many cosmological simulations or in the regions of some galaxies extending far beyond their currently known boundaries.}
{We have systematically explored parameter space to identify conditions under which plane-parallel clouds contain sufficient column densities of molecular dark matter that they could be reservoirs for future star formation. Such clouds would be undetected or at least appear by current observational criteria to be uninteresting from the perspective of star formation.}
{We use a time-dependent photon-dominated region code to produce theoretical models of the chemistry and emission arising in clouds for our chosen region of parameter space. We then select a subset of model clouds with levels of emission that are low enough to be undetectable or at least overlooked by current surveys.}
{The existence of significant column densities of cold molecular dark matter requires that the background radiation field be several or more orders of magnitude weaker than that in the solar neighbourhood. Lower turbulent velocities and cosmic ray induced ionization rates than typically associated with molecular material within a kiloparsec of the Sun are also required for the molecular matter to be dark.}
{We find that there is a large region within the parameter space that results in clouds that might contain a significant mass of molecular gas whilst remaining effectively undetectable or at least not particularly noticeable in surveys. We note briefly conditions under which molecular dark matter may contain a dynamically interesting mass.}


\maketitle


\section{Introduction}\label{Introduction}
The mapping of external galaxies in the 21\,cm \ion{H}{i} \citep[e.g.][]{Noordermeer2005,Walter2005} and the 2.6\,mm CO \citep[e.g.][]{Helfer2003,Lundgren2004} lines provides information about the distributions of atomic and molecular gas in those galaxies. Many studies of these distributions are undertaken in the context of star formation. For example, one commonly addressed question concerns whether a column density of gas in excess of a local critical value leads to star formation \citep{Kennicutt1989}.

Observations of far infrared fine structure lines indicate that in metal-poor regions, like those found in some dwarf galaxies, 2.6\,mm CO observations alone lead to a significant underestimate of the mass in molecular gas \citep{Madden1997,Madden2002}. This is consistent with the results of multiline CO studies showing that the 2.6\,mm integrated antenna temperature to molecular hydrogen conversion factor in such regions is much larger than is appropriate for Galactic giant molecular clouds, and also with the known difficulty in detecting CO emissions in some dwarf and compact galaxies compared to large spiral galaxies \citep[e.g.][]{Israel2005}. The possibility that molecular gas may be present in large quantities in such regions has been postulated \citep[see, e.g.,][and references therein]{Pfenniger1994,Combes2000,Kalberla2000}. Previous authors have concentrated on clumps with extents of the order of 10\,AU and masses very much smaller than a solar mass each. Such clumps would have a covering factor (or beam filling factor) that is much smaller than unity. Their existence might have a range of observational consequences. For instance, they may be responsible for ``extreme scattering events'', causing large variations in radio signals from more distant sources \citep{Walker1998}. We focus on lower density, larger objects having a covering factor near unity.

Of course, no survey will lead to the detection of all neutral atomic or molecular gas in a galaxy. Sensitivity limits lead to mapped distributions having boundaries, even though the real gas distributions may be more extended. The purpose of the present investigation is the determination of parameters characterizing the gas and environment that could result in significant hidden quantities of gas, with column densities comparable to those often found by surveys to be associated with star formation. Such gas could provide a reservoir for eventual star formation which would in turn create conditions leading to the gas being more easily detected. In many cases, the observed regions will be sufficiently distant that individual clouds are not observed. We have assumed, in effect, that all clouds have the same total column density and that, on average, any line of sight will pass through one cloud.

Therefore, we consider the possible existence of molecular clouds that contain a significant column density of $\mathrm{H_2}$ whilst producing sufficiently little emission that they are undetectable or at least ignored by those conducting most surveys. We produce theoretical models of the chemistry and emission arising in molecular clouds and conduct a parameter space search for physical conditions that might give rise to undetectable quantities of molecular gas. In Sect.~\ref{Model} we describe the photon-dominated region (PDR) code used to calculate the chemical and emission properties of model clouds and outline the range of parameter space examined. In Sect.~\ref{Constraints} we review the sensitivity limits of current telescopes appropriate to the detection of emission (in the IR/sub-mm/mm/radio) from molecular clouds and specify the constraints we impose for a cloud to be considered undetectable. The results of our parameter space search are presented in Sect.~\ref{Results} and we summarize our findings in Sect.~\ref{Conclusion}.


\section{Cloud model}\label{Model}
We use the \textsc{ucl\_pdr} time-dependent PDR code \citep[see][for a detailed description]{Bell2006} to calculate the chemistry, thermal balance and emission strength of fine structure and molecular lines within a model cloud, for a given set of physical parameters.

A PDR is modelled as a one-dimensional semi-infinite slab, illuminated from one side by the standard Draine radiation field \citep{Draine1978} scaled by a factor $\chi$. The chemistry and thermal balance are calculated self-consistently at each depth point into the slab and at each time-step, producing chemical abundances, emission line strengths and gas temperatures as functions of depth and time. Constant thermal pressure is assumed, and the density is calculated self-consistently with the temperature at each depth- and time-step. The turbulent contribution to the pressure is neglected. Usually the temperature varies by at most a factor of a few in most models, and the assumption of a constant density would lead to similar results. Self-gravity may be important in real clouds, and lead to some density variation. However, magnetic forces may also be important and counteract the gravity to a large extent. Given the uncertainities in the real density distributions, we consider that we are justified in making the isobaric assumption.

We adopt a reduced chemical network containing 32 species and over 300 reactions, including ion-molecule, photoionization and photodissociation reactions. Freeze-out of atoms and molecules onto grains is neglected. The gas-phase elemental abundances adopted are listed in Table~\ref{Abundances} and are assumed to scale linearly with metallicity, $Z$ (see below). The species included in the network are listed in Table~\ref{Species}. All fractional abundances in this paper are quoted relative to the hydrogen nuclei number density [$n_\mathrm{H}=n(\mathrm{H}) + 2n(\mathrm{H_2})$]. The reaction rates contained in the {\small UMIST99} database \citep{LeTeuff2000} are adopted, with some modifications introduced as part of a recent PDR benchmarking effort \citep{Rollig2006}.

\begin{table}
 \caption{Elemental abundances used in the reduced chemical network (relative to the hydrogen nuclei number density, $n_\mathrm{H}$).}
 \label{Abundances}
 \begin{center}
  \begin{tabular}{l r @{$\times$} l l r @{$\times$} l}
  \hline\hline
  He & 7.50 & $10^{-2}$ & C  & 1.42 & $10^{-4}$ \\
  O  & 3.19 & $10^{-4}$ & Mg & 5.12 & $10^{-6}$ \\
  \hline
  \end{tabular}
 \end{center}
\end{table}

\begin{table}
 \caption{Species included in the reduced chemical network.}
 \label{Species}
 \begin{center}
  \begin{tabular}{l}
  \hline\hline
  H, H$^+$, H$_2$, H$_2^+$, H$_3^+$, He, He$^+$ \\
  C, C$^+$, CH, CH$^+$, CH$_2$, CH$_2^+$, CH$_3$, \\
  CH$_3^+$, CH$_4$, CH$_4^+$, CH$_5^+$, CO, CO$^+$, HCO$^+$ \\
  O, O$^+$, OH, OH$^+$, H$_2$O, H$_2$O$^+$, H$_3$O$^+$, O$_2$, O$_2^+$ \\
  Mg, Mg$^+$, e$^-$ \\
  \hline
  \end{tabular}
 \end{center}
\end{table}

Variation in the metallicity ($Z/Z_\odot$) is taken into account through the assumption that the dust and PAH number densities, their photoelectric heating rates, the formation rate of $\mathrm{H_2}$ on grains and the elemental metal abundances all scale linearly with $Z$.

The \textsc{ucl\_pdr} code calculates the integrated intensities of the [\ion{O}{i}] 63\,$\mathrm{\umu m}$ and 146\,$\mathrm{\umu m}$, [\ion{C}{i}] 610\,$\mathrm{\umu m}$ and 370\,$\mathrm{\umu m}$, and [\ion{C}{ii}] 158\,$\mathrm{\umu m}$ fine structure lines and all CO rotational lines up to $J=11\to10$. Theoretical antenna temperatures are calculated on the assumption that emission from the surface of the cloud is distributed equally over one hemisphere (2$\pi$ steradians) and are integrated over the linewidth to give integrated line intensities (in $\mathrm{K\,km\,s^{-1}}$).


\subsection{Model parameter space}\label{Params}
Two sets of initial conditions are assumed for the chemical abundances in the models: purely atomic (all hydrogen in \element[0][]{H} and all carbon in \element[0][]{C}), or purely molecular (all hydrogen in $\mathrm{H_2}$ and all carbon in CO). These represent the two extremes for the initial state of the cloud, forming either from tenuous material or from dense cores that have subsequently relaxed to a diffuse state, therefore bracketing the range of possible starting conditions. We envisage that the clouds that are initially molecular have originated from much denser material. Consequently, the outer layers of atomic hydrogen around them are narrow. Our PDR model computes the $\mathrm{H/H_2}$ transition accurately as these objects evolve. The abundances, temperature and emission properties are calculated at each time-step for a total period of 1\,Gyr in each cloud model. Models are calculated for the entire region of parameter space with both sets of initial abundances.

Some correlation between the radiation field and the cosmic ray ionization rate would be expected for extragalactic observations where a beam samples the average interstellar UV field and  cosmic ray flux incident on a cloud, as the high-mass stars that emit the bulk of the far ultraviolet radiation also become supernovae that create remnants in which cosmic rays are accelerated. Therefore, the cosmic ray ionization rate per hydrogen molecule ($\zeta$) is assumed to scale linearly with the incident radiation field strength ($\chi$\,Draine) as:
\begin{equation}
\zeta=1.3\times10^{-17}\chi\quad\mathrm{[s^{-1}]}
\end{equation}

We choose a large region of parameter space but restrict attention to low radiation intensities ($<$$10^{-3}$\,Draine). A lower limit to the radiation field is given by the metagalactic radiation field, which has been considered extensively recently by \citet{Sternberg2002}. The spectral shapes of the standard \citet{Draine1978} and the metagalactic radiation fields differ, but a comparison at 100\,nm implies that the metagalactic field is the weaker by four orders of magnitude. Of course, that field varies from point to point depending on the distances to the nearest bright galaxies. We consider a range of four orders of magnitude in metallicity and gas pressure. Table~\ref{ParameterSpace} lists the individual parameter values considered. Since regions where the radiation background is low compared to the solar neighbourhood are far from stars which provide energy to stir their environments, such regions are likely to be fairly quiescent. The level of turbulence in the clouds considered here is expected to be significantly lower than is typically found in standard nearby Galactic giant molecular clouds \citep{Williams1995}. We therefore include models with microturbulent velocities in the range 0.05--$0.5\,\mathrm{km\,s^{-1}}$. The assumed value of this velocity is important for the heating and cooling of the clouds. It is also important in our considerations of detectability. We note that an ensemble of clouds possessing a cloud-cloud velocity dispersion would produce emission lines that are broader than those from the individual clouds; \citet{Wolfire1993} included such a cloud-cloud velocity dispersion in their models. We have not, in part because regions where few supernovae occur will have small cloud-cloud velocity dispersions, and also in part to restrict the range of parameter space to a manageable level. Our neglect of a cloud-cloud velocity dispersion leads to a more restricted region of parameter space corresponding to molecular dark matter that is difficult to detect.

In total, we constructed 384 time-dependent cloud models to cover the specified parameter space. The next section contains the criteria used to determine what part of parameter space is associated with molecular dark matter.

\begin{table}
 \caption{The range of values covered in the parameter space.}
 \label{ParameterSpace}
 \begin{center} 
  \begin{tabular}{lll}
  \hline\hline
  Parameter & \multicolumn{2}{l}{Values taken for each parameter} \\
  \hline
  $t$ & $10^1$, $10^2$, \ldots, $10^8$, $10^9$ & yr \\
  $Z$ & $10^0$, $10^{-1}$, $10^{-2}$, $10^{-3}$ & $Z_\odot$ \\
  $P/k$ & $10^1$, $10^2$, $10^3$, $10^4$ & $\mathrm{cm^{-3}\,K}$ \\
  $\chi$ & $10^{-3}$, $10^{-4}$, $10^{-5}$ & Draine \\
  $v_\mathrm{turb}$ & $0.05$, $0.1$, $0.3$, $0.5$ & $\mathrm{km\,s^{-1}}$ \\
  \hline
  \end{tabular}
 \end{center}
\end{table}


\begin{figure}[!ht]
\centering
  \includegraphics[width=8.8cm]{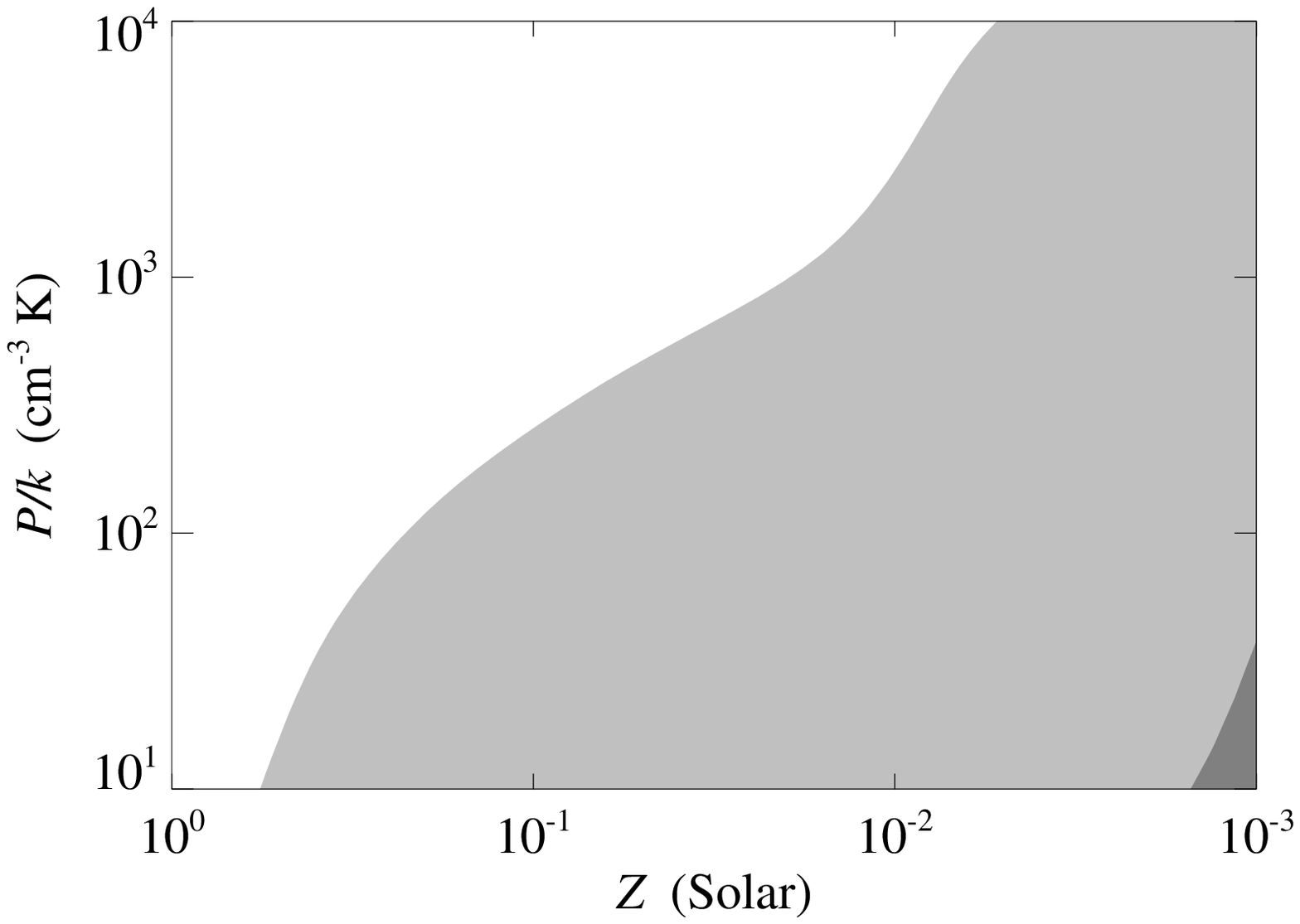}
  \includegraphics[width=8.8cm]{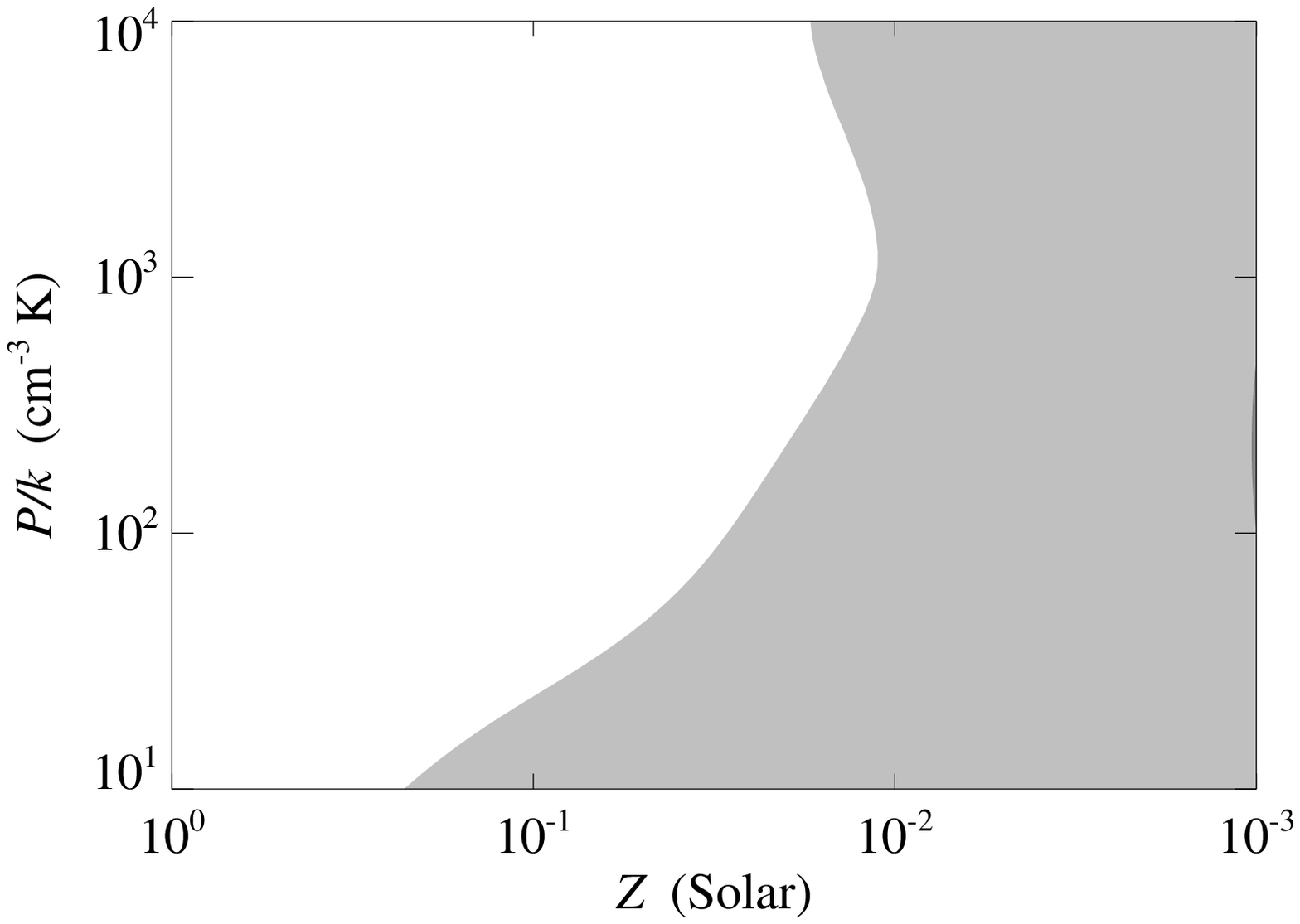}
  \includegraphics[width=8.8cm]{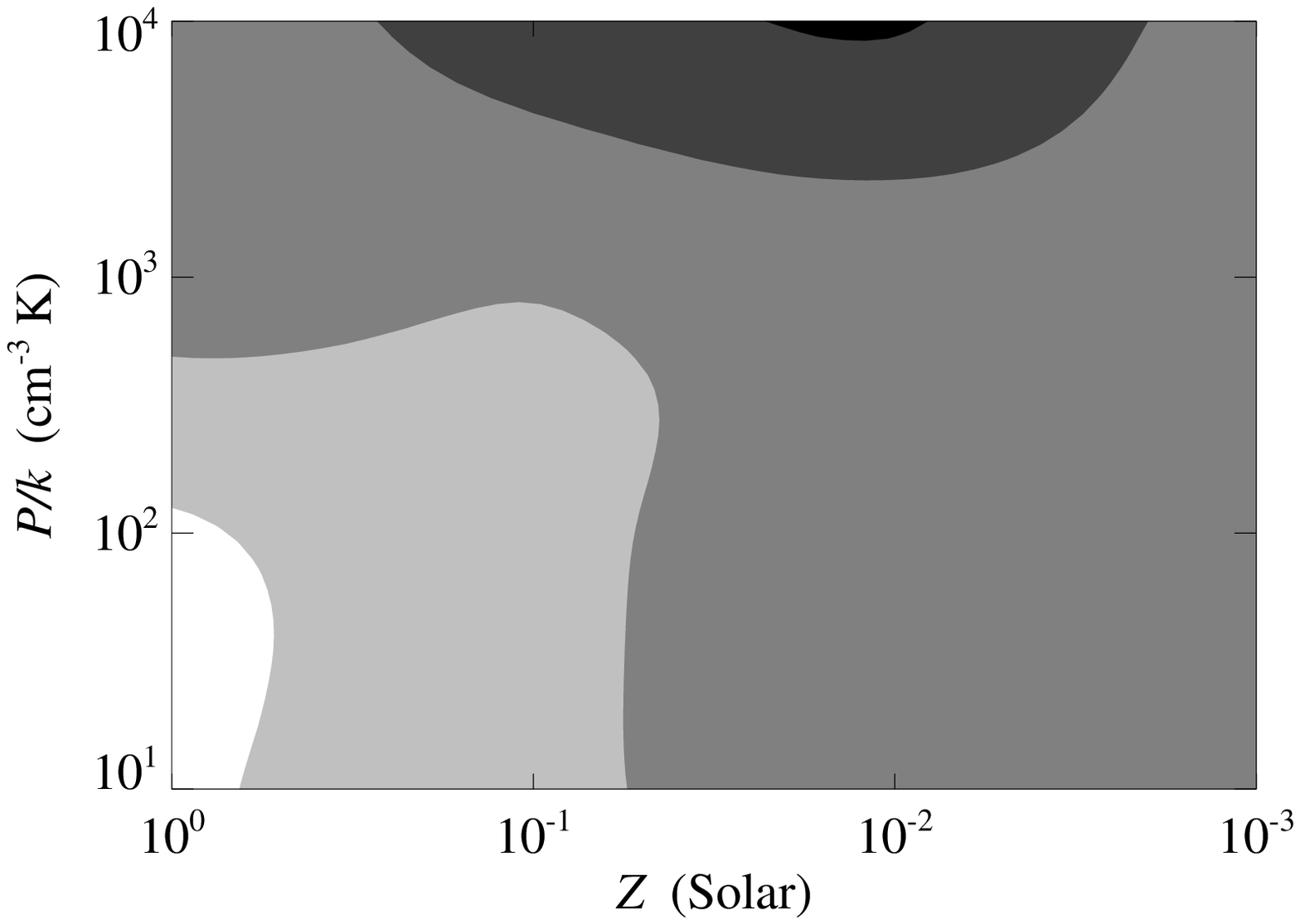}
  \caption{Contour plots of the maximum $\mathrm{H_2}$ column density produced by initially molecular cloud models with $v_\mathrm{turb}=0.05\,\mathrm{km\,s^{-1}}$ that meet the criteria of undetectable emission (see Sect.~\ref{Constraints}). Contour levels range from $N_\mathrm{H_2}<10^{21}\,\mathrm{cm^{-2}}$ (white), through $10^{21}$, $10^{22}$ and $10^{23}\,\mathrm{cm^{-2}}$, to $N_\mathrm{H_2}\ge10^{24}\,\mathrm{cm^{-2}}$ (black). The three plots show the pressure-metallicity plane of the parameter space considered, for incident radiation field strengths of $10^{-3}$, $10^{-4}$ and $10^{-5}$\,Draine, respectively.}
  \label{Plot:Radiation}
\end{figure}

\begin{table*}
 \caption{Molecular models with $v_\mathrm{turb}=0.05\,\mathrm{km\,s^{-1}}$ that satisfy the criteria of undetectable emission and $N_\mathrm{H_2}\ge10^{21}\,\mathrm{cm^{-2}}$. The value of $N_\mathrm{H_2}$ listed for each model is the peak $\mathrm{H_2}$ column density, reached at time $t_\mathrm{high}$. The period of time during which $N_\mathrm{H_2}\ge10^{21}\,\mathrm{cm^{-2}}$ in each model is defined by $t_\mathrm{min}$ and $t_\mathrm{max}$.}
 \label{Table:Molecular1}
 \begin{center}
  \input{table1}
 \end{center}
\end{table*}

\begin{figure*}
\centering
  \includegraphics[width=8.8cm]{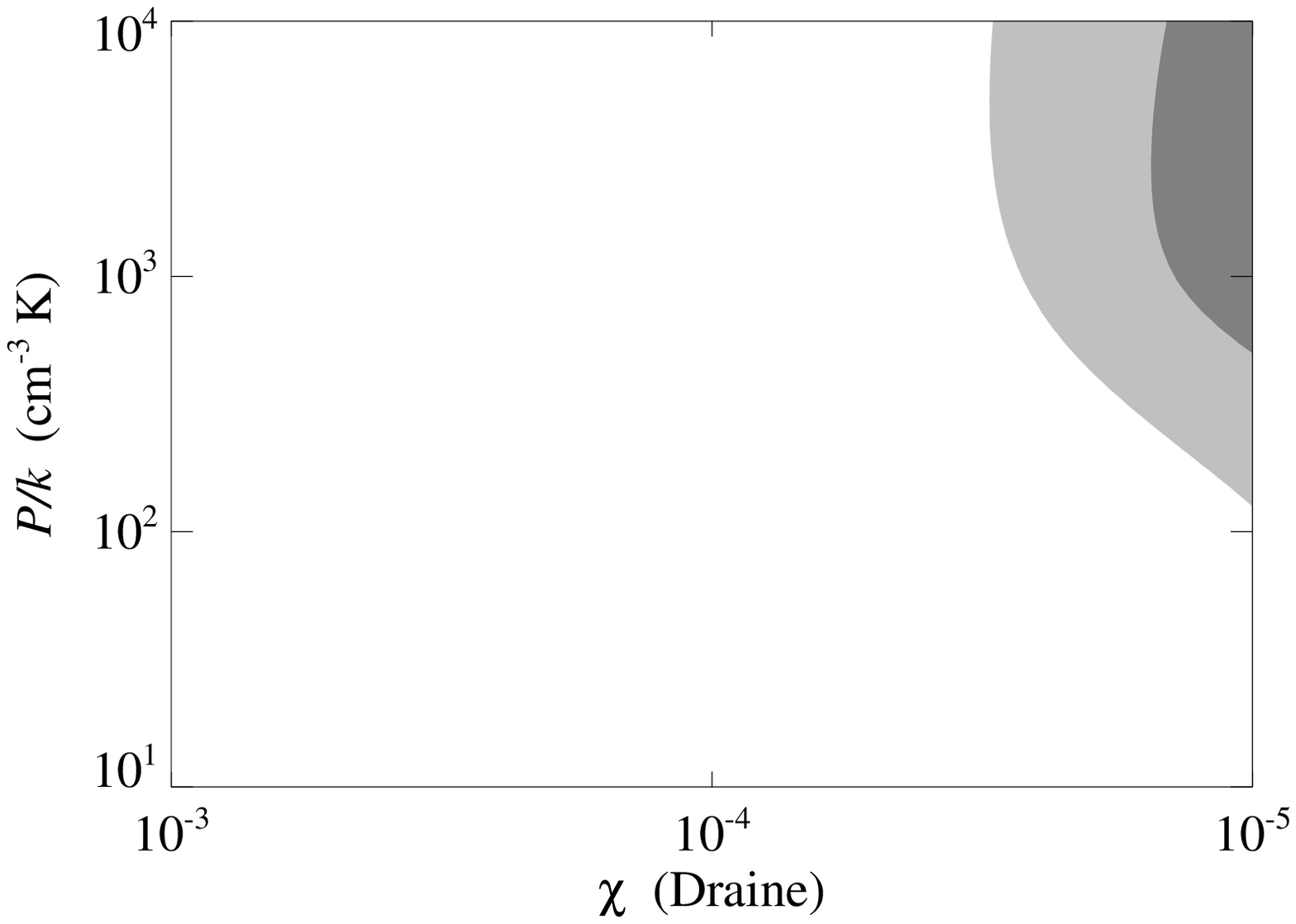}\includegraphics[width=8.8cm]{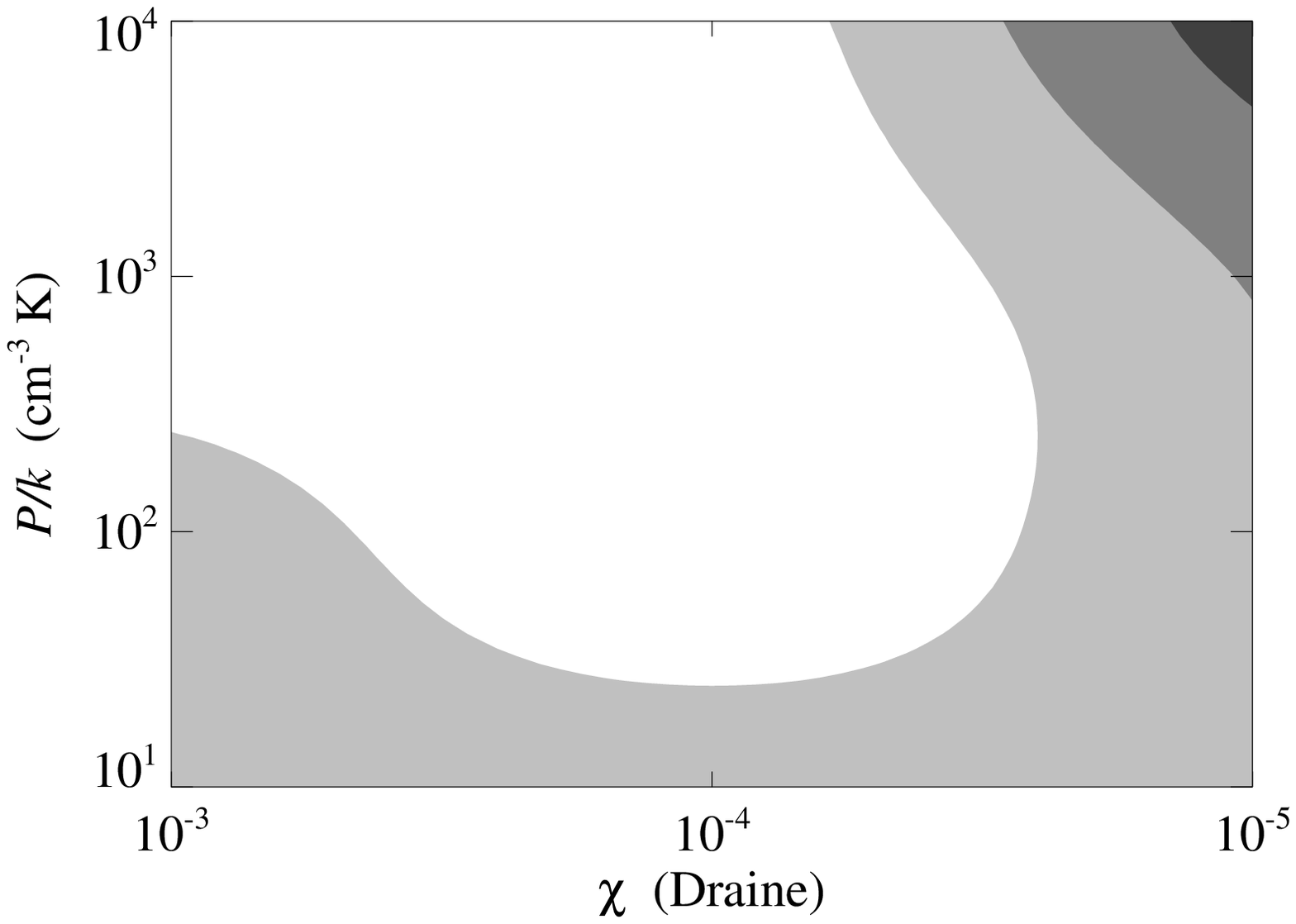}
  \includegraphics[width=8.8cm]{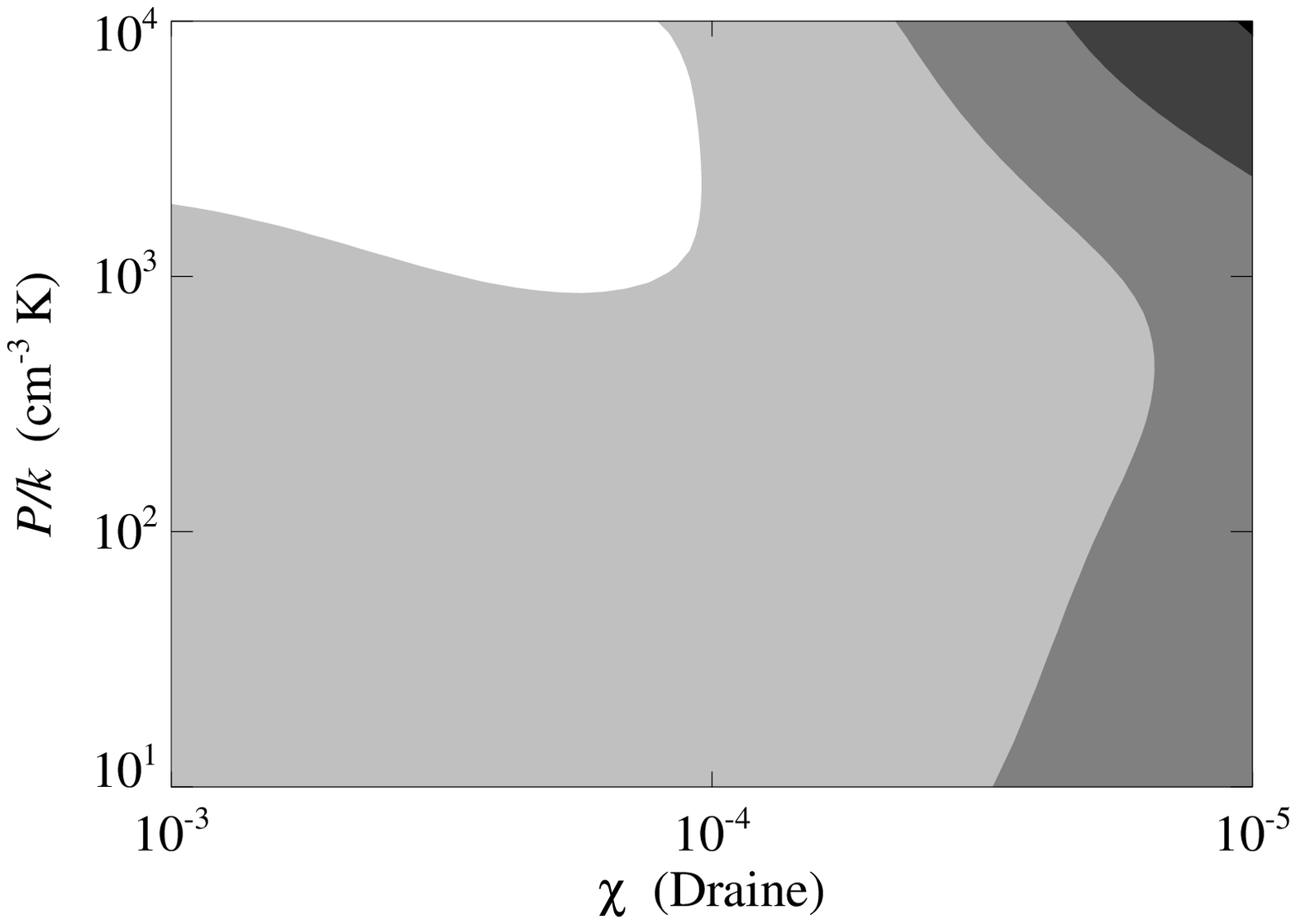}\includegraphics[width=8.8cm]{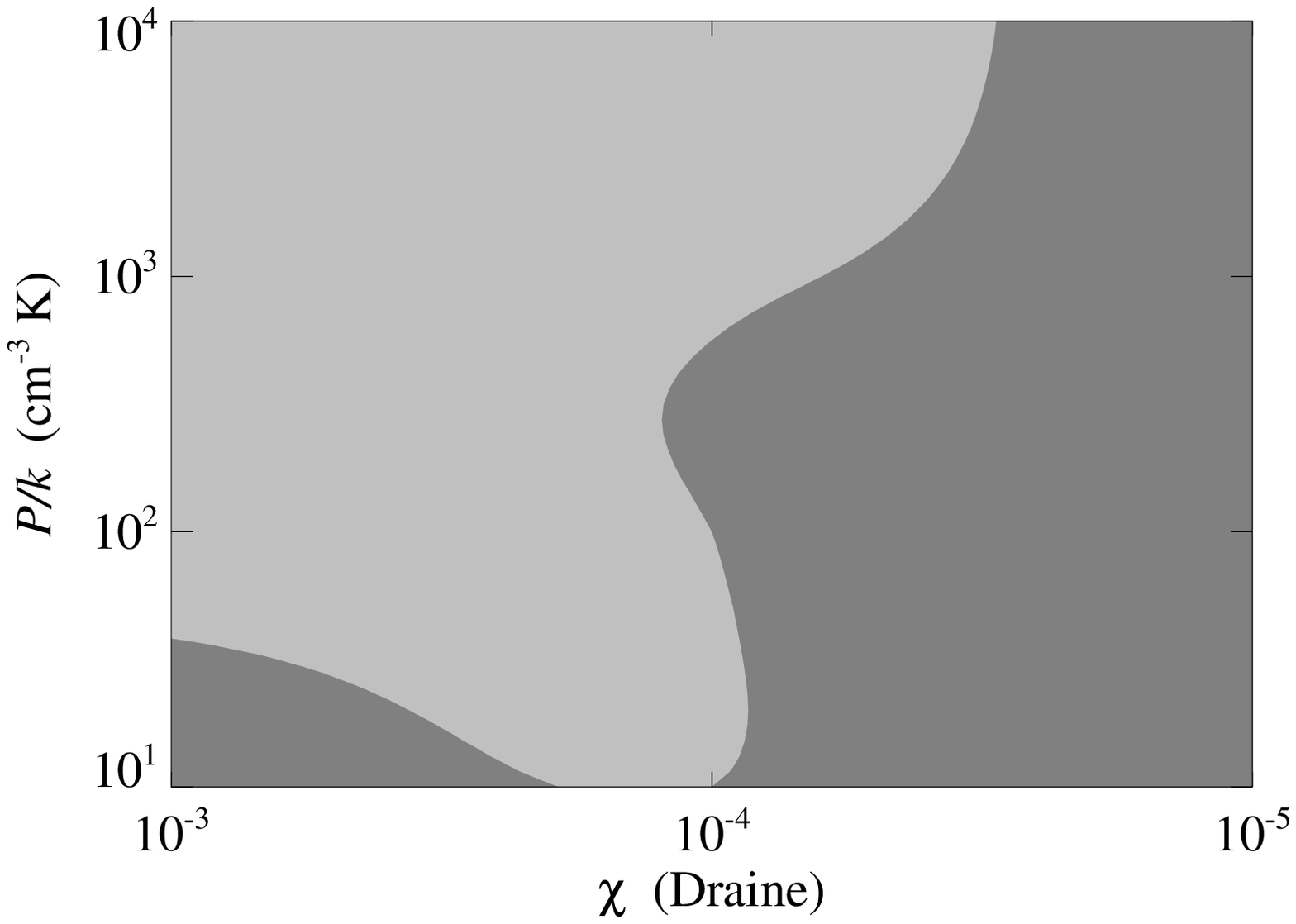}
  \caption{Contour plots of the maximum $\mathrm{H_2}$ column density produced by initially molecular cloud models with $v_\mathrm{turb}=0.05\,\mathrm{km\,s^{-1}}$ that meet the criteria of undetectable emission (see Sect.~\ref{Constraints}). Contour levels range from $N_\mathrm{H_2}<10^{21}\,\mathrm{cm^{-2}}$ (white), through $10^{21}$, $10^{22}$ and $N_\mathrm{H_2}\ge10^{23}\,\mathrm{cm^{-2}}$ (dark-grey). The four plots show the pressure-radiation plane of the parameter space considered, for metallicities of $1$, $10^{-1}$, $10^{-2}$ and $10^{-3}$ times solar (from left to right, top to bottom).}
  \label{Plot:Metallicity}
\end{figure*}


\section{Emission constraints}\label{Constraints}
The distributions of neutral atomic and molecular clouds are studied predominantly with \ion{H}{i} 21\,cm, [\ion{O}{i}], [\ion{C}{i}] and [\ion{C}{ii}] fine structure and CO rotational lines, as well as dust continuum radiation. 21\,cm surveys can lead to the six sigma detection of neutral atomic hydrogen column densities of 1.6$\times$$10^{18}\,\mathrm{cm^{-2}}$ for a velocity dispersion of $10\,\mathrm{km\,s^{-1}}$ and a comparable velocity resolution \citep[e.g.][]{Giovanelli2005}. However, published column density distributions of \ion{H}{i} in other galaxies usually do not go to such low column densities. For instance, \citet{Walter2005} take a column density of $10^{20}\,\mathrm{cm^{-2}}$ for the lowest contour in figures, and the lowest column density shown in figures by \citet{Noordermeer2005} is roughly $10^{19}\,\mathrm{cm^{-2}}$. We shall use $10^{19}\,\mathrm{cm^{-2}}$ as the minimum column density of neutral hydrogen to be considered of observational significance. Though we have cited observations that were sensitive to \ion{H}{i} column densities roughly as low as $10^{19}\,\mathrm{cm^{-2}}$, we realize that this is a rather small value. We have chosen to define such a low limiting value for the \ion{H}{i} column density because the small velocity dispersions that we have assumed would make \ion{H}{i} more readily detectable.

Ground-based observations of the lower CO rotational emission lines have a typical RMS noise limit of $\sim$$1\,\mathrm{mK}$; we therefore adopt a minimum level for the detectable integrated line intensity of $0.01\,\mathrm{K\,km\,s^{-1}}$ (assuming a velocity resolution of $10\,\mathrm{km\,s^{-1}}$), suitable for deep searches \citep[e.g.][]{Wilson1998,Smoker2000}. The noise limits on [\ion{C}{i}] 610\,$\mathrm{\umu m}$ ground-based observations are higher, but for convenience we also impose a constraint of $0.01\,\mathrm{K\,km\,s^{-1}}$ for the minimum detectable integrated line intensity.

The [\ion{O}{i}] and [\ion{C}{ii}] fine structure lines are only observable from space or the upper atmosphere. Taking typical RMS noise limits for spectra obtained with the Long Wavelength Spectrometer (LWS) on the {\em Infrared Space Observatory} ({\em ISO}), a minimum integrated flux of $\sim$$10^{-17}\,\mathrm{W\,m^{-2}}$ is reasonable (assuming a resolution of 0.29\,$\mathrm{\umu m}$ for the [\ion{O}{i}] 63\,$\mathrm{\umu m}$ line and 0.6\,$\mathrm{\umu m}$ for the [\ion{C}{ii}] 158\,$\mathrm{\umu m}$ line). This corresponds to a surface brightness of $\sim$$10^{-7}\,\mathrm{erg\,s^{-1}\,cm^{-2}\,sr^{-1}}$ \citep[assuming $\Omega_\mathrm{b}\sim10^{-7}\,\mathrm{sr}$ for the LWS;][]{Lloyd2003}. We adopt this as the detection limit for the two dominant fine structure lines.

Since the dust in these clouds is cold (typically $<$$5\,\mathrm{K}$) and depleted at the lower metallicities, its emission will be faint and therefore unlikely to be detected above the background level.

To summarize, the limits for non-detection that we impose on the cloud models are: $N_\mathrm{H\,I}<10^{19}\,\mathrm{cm^{-2}}$, $W_\mathrm{CO}<0.01\,\mathrm{K\,km\,s^{-1}}$ in all rotational lines, $W<0.01\,\mathrm{K\,km\,s^{-1}}$ for the [\ion{C}{i}] 610\,$\mathrm{\umu m}$ line, $I<10^{-7}\,\mathrm{erg\,s^{-1}\,cm^{-2}\,sr^{-1}}$ for the [\ion{O}{i}] 63\,$\mathrm{\umu m}$ and [\ion{C}{ii}] 158\,$\mathrm{\umu m}$ lines. Any cloud model with emission below all these limits simultaneously is considered to be effectively undetectable with current technology. We now describe the results of our parameter space search, presenting models that meet these criteria.


\section{Model results}\label{Results}
The column density through the cloud is restricted in each model such that the total emission in each line is below the corresponding limit given above. For each set of physical parameters, the resulting cloud model therefore represents the maximum amount of molecular gas that can be hidden under those conditions. We then select cloud models that contain significant quantities of $\mathrm{H_2}$, specifically $N_\mathrm{H_2}\ge10^{21}\,\mathrm{cm^{-2}}$, at some point during their 1\,Gyr evolution. Such a column density is of the same order of magnitude as the hydrogen nuclei column densities associated with parts of large spiral galaxies in which stars form \citep[e.g.][]{Noordermeer2005}. This subset of models defines the region of our parameter space where significant amounts of molecular gas can reside without being detected in emission. Tables~\ref{Table:Molecular1}--\ref{Table:Molecular4} list the subset of cloud models with initially molecular gas that meet our criteria for non-detectable emission whilst containing a significant amount of $\mathrm{H_2}$. Table~\ref{Table:Atomic} lists the subset of initially atomic cloud models that meet the same conditions. The tables also show the period of time for which this is the case in each model, starting at $t_\mathrm{min}$ and ending at $t_\mathrm{max}$, as well as the time, $t_\mathrm{high}$, at which the peak value of $N_\mathrm{H_2}$ is reached. In total, 111 of the 384 model clouds obtain $\mathrm{H_2}$ column densities $\ge$$10^{21}\,\mathrm{cm^{-2}}$ for some period of their evolution. The trends describing the change in total $\mathrm{H_2}$ column density with parameter variation are discussed in the next two sections.

\begin{table*}
 \caption{Molecular models with $v_\mathrm{turb}=0.1\,\mathrm{km\,s^{-1}}$ that satisfy the criteria of undetectable emission and $N_\mathrm{H_2}\ge10^{21}\,\mathrm{cm^{-2}}$. The value of $N_\mathrm{H_2}$ listed for each model is the peak $\mathrm{H_2}$ column density, reached at time $t_\mathrm{high}$. The period of time during which $N_\mathrm{H_2}\ge10^{21}\,\mathrm{cm^{-2}}$ in each model is defined by $t_\mathrm{min}$ and $t_\mathrm{max}$.}
 \label{Table:Molecular2}
 \begin{center}
  \input{table2}
 \end{center}
\end{table*}

\begin{table*}
 \caption{Molecular models with $v_\mathrm{turb}=0.3\,\mathrm{km\,s^{-1}}$ that satisfy the criteria of undetectable emission and $N_\mathrm{H_2}\ge10^{21}\,\mathrm{cm^{-2}}$. The value of $N_\mathrm{H_2}$ listed for each model is the peak $\mathrm{H_2}$ column density, reached at time $t_\mathrm{high}$. The period of time during which $N_\mathrm{H_2}\ge10^{21}\,\mathrm{cm^{-2}}$ in each model is defined by $t_\mathrm{min}$ and $t_\mathrm{max}$.}
 \label{Table:Molecular3}
 \begin{center}
  \input{table3}
 \end{center}
\end{table*}

\begin{table*}
 \caption{Molecular models with $v_\mathrm{turb}=0.5\,\mathrm{km\,s^{-1}}$ that satisfy the criteria of undetectable emission and $N_\mathrm{H_2}\ge10^{21}\,\mathrm{cm^{-2}}$. The value of $N_\mathrm{H_2}$ listed for each model is the peak $\mathrm{H_2}$ column density, reached at time $t_\mathrm{high}$. The period of time during which $N_\mathrm{H_2}\ge10^{21}\,\mathrm{cm^{-2}}$ in each model is defined by $t_\mathrm{min}$ and $t_\mathrm{max}$.}
 \label{Table:Molecular4}
 \begin{center}
  \input{table4}
 \end{center}
\end{table*}

\begin{table*}
 \caption{All atomic models that satisfy the criteria of undetectable emission and $N_\mathrm{H_2}\ge10^{21}\,\mathrm{cm^{-2}}$. The value of $N_\mathrm{H_2}$ listed for each model is the peak $\mathrm{H_2}$ column density, reached at time $t_\mathrm{high}$. The period of time during which $N_\mathrm{H_2}\ge10^{21}\,\mathrm{cm^{-2}}$ in each model is defined by $t_\mathrm{min}$ and $t_\mathrm{max}$.}
 \label{Table:Atomic}
 \begin{center}
  \input{table5}
 \end{center}
\end{table*}

\subsection{Initially molecular clouds}\label{Results:Molecular}

For models for which the gas is initially molecular, the peak $\mathrm{H_2}$ column density ($>$$10^{24}\,\mathrm{cm^{-2}}$) occurs in the cloud with the lowest turbulent velocity, subject to the lowest incident radiation field strength, with low metallicity and with the highest pressure ($0.05\,\mathrm{km\,s^{-1}}$, $10^{-5}$\,Draine, $10^{-2}\,Z_\odot$, $10^{4}\,\mathrm{cm^{-3}\,K}$; see Table~\ref{Table:Molecular1} and Fig.~\ref{Plot:Radiation}). For the lowest radiation field strength, in the models with low turbulent velocities (0.05 or $0.1\,\mathrm{km\,s^{-1}}$), the $\mathrm{H_2}$ column density increases as the pressure rises and it is the intermediate metallicities ($10^{-1}$--$10^{-2}\,Z_\odot$) that produce the highest column densities of dark material. For models for which the radiation field does not have its lowest value, the $\mathrm{H_2}$ column density increases as the pressure and metallicity drop (see Fig.~\ref{Plot:Radiation}).

At the lowest metallicity considered ($10^{-3}\,Z_\odot$), the $\mathrm{H_2}$ column density is $\ge$$10^{21}\,\mathrm{cm^{-2}}$ for almost all combinations of turbulent velocity, radiation field strength and pressure (see the bottom right panel of Fig.~\ref{Plot:Metallicity} for the $v_\mathrm{turb}=0.05\,\mathrm{km\,s^{-1}}$ case). The only exceptions are the three cloud models with the lowest incident radiation and highest pressure at turbulent velocities of $0.3$ and $0.5\,\mathrm{km\,s^{-1}}$ (see Tables~\ref{Table:Molecular3} and \ref{Table:Molecular4}). Overall, increasing turbulent velocity within the cloud serves to reduce the $\mathrm{H_2}$ column density that can be hidden, although some of the low pressure models show a slight increase in $N_\mathrm{H_2}$ at higher turbulent velocities.

The period of time for which $N_\mathrm{H_2}\ge10^{21}\,\mathrm{cm^{-2}}$ in each model (i.e.~from $t_\mathrm{min}$ to $t_\mathrm{max}$) is governed by the restrictions placed on the CO and \ion{H}{i} emission. For cloud models with $t_\mathrm{min}>0$, the 2.6\,mm CO line emission is initially above the limit given in Sect.~\ref{Constraints} and therefore considered detectable, but falls below the limit once $t_\mathrm{min}$ is reached. In models with $t_\mathrm{max}<1$\,Gyr, the maximum time considered in the calculations, the 21\,cm \ion{H}{i} emission rises above its detection limit at late times, reducing the period for which the cloud remains undetectable. This is generally the case for models with higher radiation field strengths.

An analysis of the thermal balance in the models shows that the dominant cooling processes are emission in the lower rotational transitions of CO and the fine structure lines of neutral carbon, with some contribution from collisions with the cooler dust grains in the high pressure models. The heating of the gas at the cloud surface is dominated by $\mathrm{H_2}$ formation and photodissociation, whilst turbulence and cosmic ray ionization become important heating mechanisms deeper into the cloud. In models with higher radiation field strengths, PAH photoelectric heating plays a role near the cloud surface. Heating due to turbulent dissipation becomes the dominant heating process at higher turbulent velocities.

\subsection{Initially atomic clouds}\label{Results:Atomic}
From Table~\ref{Table:Atomic} it can be seen that there are significantly fewer models for which the gas is initially atomic that can produce appreciable quantities of hidden molecular gas. This is mainly due to the relatively long time-scales required for molecular hydrogen formation under the physical conditions being considered here. However, despite this limitation, there still exists a well-defined region of parameter space within which atomic gas can evolve to form large column densities of molecular gas whilst remaining undetectable in emission by the usual tracers.

In all the models listed in Table~\ref{Table:Atomic}, the radiation field strength is the lowest considered in our parameter space ($10^{-5}$\,Draine) and the pressures are among the highest ($10^{3}$ or $10^{4}\,\mathrm{cm^{-3}\,K}$). There is some spread in metallicity, but there are no clouds with turbulent velocities above $0.1\,\mathrm{km\,s^{-1}}$ that have $\ge$$10^{21}\,\mathrm{cm^{-2}}$ of $\mathrm{H_2}$. The peak $\mathrm{H_2}$ column density is 7.4$\times$$10^{22}\,\mathrm{cm^{-2}}$, some 20 times smaller than the largest column density obtained by clouds evolving from molecular gas. None of the initially atomic clouds reach $N_\mathrm{H_2}\ge10^{21}\,\mathrm{cm^{-2}}$ until they have existed for 20\,Myr or longer; several of the cloud models do not obtain significant $\mathrm{H_2}$ column densities until 1\,Gyr has passed. The heating and cooling of the gas is governed by the processes discussed in the previous section, although the heating due to the formation of $\mathrm{H_2}$ is more significant in these models, since the hydrogen is initially atomic.


\section{Conclusion}\label{Conclusion}
We have conducted a large study of the parameter space associated with regions where the radiation field is weak relative to that in the solar vicinity to search for conditions that give rise to significant quantities of molecular gas with emission strengths that are comparable to the lower limits considered in standard surveys. Having explored parameter space to identify the time-dependent cloud models that satisfy these limits, we selected those models in which a significant column density of molecular hydrogen, $N_\mathrm{H_2}\ge10^{21}\,\mathrm{cm^{-2}}$ was reached.

We find that there is a large region within the parameter space that meets these requirements, resulting in clouds that might contain a significant mass of molecular gas whilst remaining effectively undetectable or at least not particularly noticeable in surveys. This region is characterized by low turbulent velocities ($\le$$0.1\,\mathrm{km\,s^{-1}}$), low radiation fields ($\le$$10^{-4}$\,Draine), normal to low metallicities ($\sim$1--$10^{-2}\,Z_\odot$) and intermediate pressures ($\sim$$10^{3}$--$10^{4}\,\mathrm{cm^{-3}\,K}$). Clouds forming from molecular gas under these conditions are capable of maintaining large amounts of $\mathrm{H_2}$ without becoming detectable in emission. Initially atomic gas can also form significant column densities of molecular material with undetectable emission, but the time-scales required are much longer ($>$$10^{7}$\,yr).

\citet{Sternberg2002} conducted a study of PDRs in which they assumed the present metagalactic radiation field to be incident on clouds in work on minihaloes composed of cold dark matter of the variety considered in many cosmological studies and on compact high velocity \ion{H}{i} clouds. As mentioned above, the radiation field that they assumed is of comparable strength at 100\,nm to fields for which we found large amounts of dark molecular gas to exist. Hence, there is some overlap between their work and ours. However, their emphasis is on \ion{H}{i} clouds detectable in 21\,cm emission, whereas ours is on the possible existence of dark molecular gas with substantial enough column densities to serve as reservoirs for potential star formation. They address observations of \ion{H}{i} clouds having total hydrogen nuclei column densities of an order of magnitude and more below those with which we are concerned. Our interest is in whether regions of much higher column density may harbour sufficient molecular gas to become sites of stellar birth while they remain hidden in current surveys. We also wonder whether molecular dark matter existing in regions far away from bright stars may lead to future star formation in some galaxies occurring significantly beyond the currently detected bounds of those galaxies.

Finally, we note that an $\mathrm{H_2}$ column density of $10^{21}\,\mathrm{cm^{-2}}$ distributed evenly over a disc with a radius of 40\,kpc would correspond to a mass of $10^{11}\,M_\odot$, sufficient to be of dynamical importance in some galaxies.

A means of placing limits on the product of the covering factor of the type of molecular dark matter we are considering and the mean surface area per galaxy over which such dark matter is distributed would be through millimetre-wave absorption observations of other galaxies similar to those made by \citet{Liszt1998} to study diffuse clouds in our galaxy. Observations towards many background sources would almost certainly be required to set meaningful limits on that product, as a detection rate of about 4 optical QSO absorption systems per unit redshift at a redshift of $\sim$1 implies that intervening galaxies must have radii of almost 100\,kpc if they are randomly oriented discs \citep[e.g.][]{Sargent1979}. \citet{Zwaan2006} have argued that a redshift interval of $\Delta z\approx100$ has been covered in the CORALS survey of radio-selected quasars \citep{Ellison2001}, whereas a $\Delta z$ range on the order of 3300 near $z=0$ should be required to discover a high $\mathrm{H_2}$ column density absorber at the present epoch, if $\mathrm{H_2}$ has the sort of distributions in galaxies that are inferred from the BIMA SONG maps of CO in nearby galaxies \citep{Helfer2003}.


\begin{acknowledgements}
TAB is supported by a PPARC studentship. SV acknowledges individual financial support from a PPARC Advanced Fellowship. We thank Professor Peter Brand and Dr.~Jonathan Rawlings for helpful discussions. We also thank the referee and the editor, Professor Malcolm Walmsley, for useful comments which helped to improve an earlier draft of this paper.
\end{acknowledgements}


\end{document}

%% file: table1.tex
\begin{tabular}{r@{}l c c r@{$\times$}l r@{$\times$}l r@{$\times$}l r@{$\times$}l}
 \hline\hline
 \multicolumn{2}{c}{$Z$} & $\chi$ & $P/k$ & \multicolumn{2}{c}{$N_{\rm H_2}$} & \multicolumn{2}{c}{$t_{\rm high}$} & \multicolumn{2}{c}{$t_{\rm min}$} & \multicolumn{2}{c}{$t_{\rm max}$} \\
 \multicolumn{2}{c}{($Z_\odot$)} & (Draine) & (cm$^{-3}$ K) & \multicolumn{2}{c}{(cm$^{-2}$)} & \multicolumn{2}{c}{(yr)} & \multicolumn{2}{c}{(yr)} & \multicolumn{2}{c}{(yr)} \\
 \hline
 10&$^{0}$  & 10$^{-5}$ & 10$^{3}$ & 2.9&10$^{22}$ & 7&10$^{6}$ & 0&10$^{0}$ & 1&10$^{9}$ \\
 10&$^{0}$  & 10$^{-5}$ & 10$^{4}$ & 2.9&10$^{22}$ & 3&10$^{7}$ & 0&10$^{0}$ & 1&10$^{9}$ \\
 \hline
 10&$^{-1}$ & 10$^{-3}$ & 10$^{1}$ & 2.4&10$^{21}$ & 1&10$^{7}$ & 6&10$^{6}$ & 1&10$^{7}$ \\
 10&$^{-1}$ & 10$^{-3}$ & 10$^{2}$ & 2.2&10$^{21}$ & 4&10$^{7}$ & 7&10$^{6}$ & 1&10$^{8}$ \\
 10&$^{-1}$ & 10$^{-4}$ & 10$^{1}$ & 2.3&10$^{21}$ & 6&10$^{8}$ & 8&10$^{7}$ & 8&10$^{8}$ \\
 10&$^{-1}$ & 10$^{-5}$ & 10$^{1}$ & 6.0&10$^{21}$ & 3&10$^{7}$ & 0&10$^{0}$ & 1&10$^{9}$ \\
 10&$^{-1}$ & 10$^{-5}$ & 10$^{2}$ & 6.3&10$^{21}$ & 4&10$^{3}$ & 0&10$^{0}$ & 1&10$^{9}$ \\
 10&$^{-1}$ & 10$^{-5}$ & 10$^{3}$ & 1.2&10$^{22}$ & 1&10$^{6}$ & 0&10$^{0}$ & 1&10$^{9}$ \\
 10&$^{-1}$ & 10$^{-5}$ & 10$^{4}$ & 2.8&10$^{23}$ & 4&10$^{7}$ & 0&10$^{0}$ & 1&10$^{9}$ \\
 \hline
 10&$^{-2}$ & 10$^{-3}$ & 10$^{1}$ & 5.3&10$^{21}$ & 9&10$^{6}$ & 3&10$^{6}$ & 1&10$^{7}$ \\
 10&$^{-2}$ & 10$^{-3}$ & 10$^{2}$ & 4.7&10$^{21}$ & 8&10$^{6}$ & 4&10$^{6}$ & 1&10$^{7}$ \\
 10&$^{-2}$ & 10$^{-3}$ & 10$^{3}$ & 1.6&10$^{21}$ & 5&10$^{7}$ & 2&10$^{7}$ & 1&10$^{8}$ \\
 10&$^{-2}$ & 10$^{-4}$ & 10$^{1}$ & 5.5&10$^{21}$ & 9&10$^{7}$ & 0&10$^{0}$ & 1&10$^{8}$ \\
 10&$^{-2}$ & 10$^{-4}$ & 10$^{2}$ & 2.8&10$^{21}$ & 5&10$^{8}$ & 0&10$^{0}$ & 1&10$^{9}$ \\
 10&$^{-2}$ & 10$^{-4}$ & 10$^{3}$ & 1.1&10$^{21}$ & 1&10$^{9}$ & 0&10$^{0}$ & 1&10$^{9}$ \\
 10&$^{-2}$ & 10$^{-4}$ & 10$^{4}$ & 1.4&10$^{21}$ & 5&10$^{3}$ & 0&10$^{0}$ & 1&10$^{7}$ \\
 10&$^{-2}$ & 10$^{-5}$ & 10$^{1}$ & 2.6&10$^{22}$ & 0&10$^{0}$ & 0&10$^{0}$ & 1&10$^{9}$ \\
 10&$^{-2}$ & 10$^{-5}$ & 10$^{2}$ & 2.4&10$^{22}$ & 3&10$^{5}$ & 0&10$^{0}$ & 1&10$^{9}$ \\
 10&$^{-2}$ & 10$^{-5}$ & 10$^{3}$ & 2.8&10$^{22}$ & 0&10$^{0}$ & 0&10$^{0}$ & 1&10$^{9}$ \\
 10&$^{-2}$ & 10$^{-5}$ & 10$^{4}$ & 1.2&10$^{24}$ & 0&10$^{0}$ & 0&10$^{0}$ & 1&10$^{9}$ \\
 \hline
 10&$^{-3}$ & 10$^{-3}$ & 10$^{1}$ & 1.2&10$^{22}$ & 9&10$^{6}$ & 2&10$^{5}$ & 1&10$^{7}$ \\
 10&$^{-3}$ & 10$^{-3}$ & 10$^{2}$ & 9.1&10$^{21}$ & 1&10$^{7}$ & 2&10$^{6}$ & 1&10$^{7}$ \\
 10&$^{-3}$ & 10$^{-3}$ & 10$^{3}$ & 7.9&10$^{21}$ & 9&10$^{6}$ & 3&10$^{6}$ & 1&10$^{7}$ \\
 10&$^{-3}$ & 10$^{-3}$ & 10$^{4}$ & 3.5&10$^{21}$ & 2&10$^{7}$ & 3&10$^{6}$ & 7&10$^{7}$ \\
 10&$^{-3}$ & 10$^{-4}$ & 10$^{1}$ & 1.0&10$^{22}$ & 8&10$^{7}$ & 0&10$^{0}$ & 1&10$^{8}$ \\
 10&$^{-3}$ & 10$^{-4}$ & 10$^{2}$ & 1.0&10$^{22}$ & 9&10$^{7}$ & 0&10$^{0}$ & 1&10$^{8}$ \\
 10&$^{-3}$ & 10$^{-4}$ & 10$^{3}$ & 9.1&10$^{21}$ & 2&10$^{8}$ & 0&10$^{0}$ & 9&10$^{8}$ \\
 10&$^{-3}$ & 10$^{-4}$ & 10$^{4}$ & 3.3&10$^{21}$ & 1&10$^{9}$ & 0&10$^{0}$ & 1&10$^{9}$ \\
 10&$^{-3}$ & 10$^{-5}$ & 10$^{1}$ & 2.1&10$^{22}$ & 9&10$^{8}$ & 0&10$^{0}$ & 1&10$^{9}$ \\
 10&$^{-3}$ & 10$^{-5}$ & 10$^{2}$ & 2.0&10$^{22}$ & 9&10$^{4}$ & 0&10$^{0}$ & 1&10$^{9}$ \\
 10&$^{-3}$ & 10$^{-5}$ & 10$^{3}$ & 2.0&10$^{22}$ & 2&10$^{4}$ & 0&10$^{0}$ & 1&10$^{9}$ \\
 10&$^{-3}$ & 10$^{-5}$ & 10$^{4}$ & 3.2&10$^{22}$ & 2&10$^{5}$ & 0&10$^{0}$ & 1&10$^{9}$ \\
 \hline
\end{tabular}

%% file: table2.tex
\begin{tabular}{r@{}l c c r@{$\times$}l r@{$\times$}l r@{$\times$}l r@{$\times$}l}
 \hline\hline
 \multicolumn{2}{c}{$Z$} & $\chi$ & $P/k$ & \multicolumn{2}{c}{$N_{\rm H_2}$} & \multicolumn{2}{c}{$t_{\rm high}$} & \multicolumn{2}{c}{$t_{\rm min}$} & \multicolumn{2}{c}{$t_{\rm max}$} \\
 \multicolumn{2}{c}{($Z_\odot$)} & (Draine) & (cm$^{-3}$ K) & \multicolumn{2}{c}{(cm$^{-2}$)} & \multicolumn{2}{c}{(yr)} & \multicolumn{2}{c}{(yr)} & \multicolumn{2}{c}{(yr)} \\
 \hline
 10&$^{0}$  & 10$^{-5}$ & 10$^{4}$ & 2.9&10$^{22}$ & 2&10$^{6}$ & 0&10$^{0}$ & 1&10$^{9}$ \\
 \hline
 10&$^{-1}$ & 10$^{-3}$ & 10$^{1}$ & 1.4&10$^{21}$ & 9&10$^{6}$ & 8&10$^{6}$ & 1&10$^{7}$ \\
 10&$^{-1}$ & 10$^{-3}$ & 10$^{2}$ & 2.6&10$^{21}$ & 4&10$^{7}$ & 1&10$^{7}$ & 1&10$^{8}$ \\
 10&$^{-1}$ & 10$^{-4}$ & 10$^{1}$ & 2.5&10$^{21}$ & 5&10$^{8}$ & 1&10$^{8}$ & 1&10$^{9}$ \\
 10&$^{-1}$ & 10$^{-5}$ & 10$^{1}$ & 2.7&10$^{21}$ & 6&10$^{5}$ & 0&10$^{0}$ & 1&10$^{9}$ \\
 10&$^{-1}$ & 10$^{-5}$ & 10$^{2}$ & 2.7&10$^{21}$ & 0&10$^{0}$ & 0&10$^{0}$ & 1&10$^{9}$ \\
 10&$^{-1}$ & 10$^{-5}$ & 10$^{3}$ & 3.1&10$^{21}$ & 2&10$^{5}$ & 0&10$^{0}$ & 1&10$^{9}$ \\
 10&$^{-1}$ & 10$^{-5}$ & 10$^{4}$ & 7.0&10$^{22}$ & 5&10$^{5}$ & 0&10$^{0}$ & 1&10$^{9}$ \\
 \hline
 10&$^{-2}$ & 10$^{-3}$ & 10$^{1}$ & 5.0&10$^{21}$ & 9&10$^{6}$ & 4&10$^{6}$ & 1&10$^{7}$ \\
 10&$^{-2}$ & 10$^{-3}$ & 10$^{2}$ & 4.0&10$^{21}$ & 8&10$^{6}$ & 4&10$^{6}$ & 1&10$^{7}$ \\
 10&$^{-2}$ & 10$^{-3}$ & 10$^{3}$ & 1.7&10$^{21}$ & 7&10$^{7}$ & 2&10$^{7}$ & 9&10$^{7}$ \\
 10&$^{-2}$ & 10$^{-4}$ & 10$^{1}$ & 4.5&10$^{21}$ & 8&10$^{7}$ & 4&10$^{7}$ & 1&10$^{8}$ \\
 10&$^{-2}$ & 10$^{-4}$ & 10$^{2}$ & 3.0&10$^{21}$ & 4&10$^{8}$ & 6&10$^{7}$ & 1&10$^{9}$ \\
 10&$^{-2}$ & 10$^{-4}$ & 10$^{3}$ & 1.0&10$^{21}$ & 9&10$^{8}$ & 8&10$^{8}$ & 9&10$^{8}$ \\
 10&$^{-2}$ & 10$^{-5}$ & 10$^{1}$ & 3.0&10$^{21}$ & 9&10$^{3}$ & 0&10$^{0}$ & 1&10$^{9}$ \\
 10&$^{-2}$ & 10$^{-5}$ & 10$^{2}$ & 2.9&10$^{21}$ & 0&10$^{0}$ & 0&10$^{0}$ & 1&10$^{9}$ \\
 10&$^{-2}$ & 10$^{-5}$ & 10$^{3}$ & 3.1&10$^{21}$ & 2&10$^{6}$ & 0&10$^{0}$ & 1&10$^{9}$ \\
 10&$^{-2}$ & 10$^{-5}$ & 10$^{4}$ & 4.6&10$^{21}$ & 7&10$^{5}$ & 0&10$^{0}$ & 1&10$^{9}$ \\
 \hline
 10&$^{-3}$ & 10$^{-3}$ & 10$^{1}$ & 1.2&10$^{22}$ & 1&10$^{7}$ & 5&10$^{4}$ & 1&10$^{7}$ \\
 10&$^{-3}$ & 10$^{-3}$ & 10$^{2}$ & 1.1&10$^{22}$ & 1&10$^{7}$ & 2&10$^{6}$ & 1&10$^{7}$ \\
 10&$^{-3}$ & 10$^{-3}$ & 10$^{3}$ & 6.8&10$^{21}$ & 9&10$^{6}$ & 3&10$^{6}$ & 1&10$^{7}$ \\
 10&$^{-3}$ & 10$^{-3}$ & 10$^{4}$ & 4.0&10$^{21}$ & 3&10$^{7}$ & 5&10$^{6}$ & 8&10$^{7}$ \\
 10&$^{-3}$ & 10$^{-4}$ & 10$^{1}$ & 1.1&10$^{22}$ & 1&10$^{8}$ & 2&10$^{6}$ & 1&10$^{8}$ \\
 10&$^{-3}$ & 10$^{-4}$ & 10$^{2}$ & 1.0&10$^{22}$ & 9&10$^{7}$ & 8&10$^{6}$ & 1&10$^{8}$ \\
 10&$^{-3}$ & 10$^{-4}$ & 10$^{3}$ & 7.8&10$^{21}$ & 2&10$^{8}$ & 2&10$^{7}$ & 1&10$^{9}$ \\
 10&$^{-3}$ & 10$^{-4}$ & 10$^{4}$ & 3.0&10$^{21}$ & 9&10$^{8}$ & 5&10$^{7}$ & 1&10$^{9}$ \\
 10&$^{-3}$ & 10$^{-5}$ & 10$^{1}$ & 1.1&10$^{22}$ & 9&10$^{8}$ & 0&10$^{0}$ & 1&10$^{9}$ \\
 10&$^{-3}$ & 10$^{-5}$ & 10$^{2}$ & 8.1&10$^{21}$ & 1&10$^{9}$ & 0&10$^{0}$ & 1&10$^{9}$ \\
 10&$^{-3}$ & 10$^{-5}$ & 10$^{3}$ & 3.1&10$^{21}$ & 1&10$^{9}$ & 0&10$^{0}$ & 1&10$^{9}$ \\
 10&$^{-3}$ & 10$^{-5}$ & 10$^{4}$ & 1.7&10$^{21}$ & 4&10$^{5}$ & 0&10$^{0}$ & 1&10$^{9}$ \\
 \hline
\end{tabular}

%% file: table3.tex
\begin{tabular}{r@{}l c c r@{$\times$}l r@{$\times$}l r@{$\times$}l r@{$\times$}l}
 \hline\hline
 \multicolumn{2}{c}{$Z$} & $\chi$ & $P/k$ & \multicolumn{2}{c}{$N_{\rm H_2}$} & \multicolumn{2}{c}{$t_{\rm high}$} & \multicolumn{2}{c}{$t_{\rm min}$} & \multicolumn{2}{c}{$t_{\rm max}$} \\
 \multicolumn{2}{c}{($Z_\odot$)} & (Draine) & (cm$^{-3}$ K) & \multicolumn{2}{c}{(cm$^{-2}$)} & \multicolumn{2}{c}{(yr)} & \multicolumn{2}{c}{(yr)} & \multicolumn{2}{c}{(yr)} \\
 \hline
 10&$^{-1}$ & 10$^{-3}$ & 10$^{1}$ & 2.0&10$^{21}$ & 5&10$^{6}$ & 5&10$^{6}$ & 1&10$^{7}$ \\
 10&$^{-1}$ & 10$^{-3}$ & 10$^{2}$ & 2.3&10$^{21}$ & 1&10$^{8}$ & 7&10$^{6}$ & 1&10$^{8}$ \\
 10&$^{-1}$ & 10$^{-4}$ & 10$^{1}$ & 3.4&10$^{21}$ & 9&10$^{8}$ & 4&10$^{7}$ & 1&10$^{9}$ \\
 \hline
 10&$^{-2}$ & 10$^{-3}$ & 10$^{1}$ & 5.6&10$^{21}$ & 9&10$^{6}$ & 3&10$^{6}$ & 1&10$^{7}$ \\
 10&$^{-2}$ & 10$^{-3}$ & 10$^{2}$ & 4.4&10$^{21}$ & 9&10$^{6}$ & 3&10$^{6}$ & 1&10$^{7}$ \\
 10&$^{-2}$ & 10$^{-3}$ & 10$^{3}$ & 1.5&10$^{21}$ & 3&10$^{7}$ & 5&10$^{6}$ & 1&10$^{8}$ \\
 10&$^{-2}$ & 10$^{-4}$ & 10$^{1}$ & 5.1&10$^{21}$ & 7&10$^{7}$ & 3&10$^{7}$ & 1&10$^{8}$ \\
 10&$^{-2}$ & 10$^{-4}$ & 10$^{2}$ & 5.5&10$^{21}$ & 2&10$^{8}$ & 4&10$^{7}$ & 1&10$^{9}$ \\
 10&$^{-2}$ & 10$^{-5}$ & 10$^{1}$ & 5.3&10$^{21}$ & 1&10$^{9}$ & 4&10$^{8}$ & 1&10$^{9}$ \\
 10&$^{-2}$ & 10$^{-5}$ & 10$^{2}$ & 1.7&10$^{21}$ & 1&10$^{9}$ & 5&10$^{8}$ & 1&10$^{9}$ \\
 \hline
 10&$^{-3}$ & 10$^{-3}$ & 10$^{1}$ & 1.2&10$^{22}$ & 9&10$^{6}$ & 0&10$^{0}$ & 1&10$^{7}$ \\
 10&$^{-3}$ & 10$^{-3}$ & 10$^{2}$ & 1.2&10$^{22}$ & 1&10$^{7}$ & 2&10$^{5}$ & 1&10$^{7}$ \\
 10&$^{-3}$ & 10$^{-3}$ & 10$^{3}$ & 6.0&10$^{21}$ & 9&10$^{6}$ & 3&10$^{6}$ & 1&10$^{7}$ \\
 10&$^{-3}$ & 10$^{-3}$ & 10$^{4}$ & 1.9&10$^{21}$ & 7&10$^{6}$ & 5&10$^{6}$ & 1&10$^{8}$ \\
 10&$^{-3}$ & 10$^{-4}$ & 10$^{1}$ & 1.3&10$^{22}$ & 1&10$^{8}$ & 0&10$^{0}$ & 1&10$^{8}$ \\
 10&$^{-3}$ & 10$^{-4}$ & 10$^{2}$ & 9.3&10$^{21}$ & 9&10$^{7}$ & 2&10$^{6}$ & 1&10$^{8}$ \\
 10&$^{-3}$ & 10$^{-4}$ & 10$^{3}$ & 3.9&10$^{21}$ & 7&10$^{7}$ & 3&10$^{7}$ & 1&10$^{9}$ \\
 10&$^{-3}$ & 10$^{-4}$ & 10$^{4}$ & 2.2&10$^{21}$ & 6&10$^{8}$ & 8&10$^{7}$ & 1&10$^{9}$ \\
 10&$^{-3}$ & 10$^{-5}$ & 10$^{1}$ & 1.2&10$^{22}$ & 1&10$^{9}$ & 0&10$^{0}$ & 1&10$^{9}$ \\
 10&$^{-3}$ & 10$^{-5}$ & 10$^{2}$ & 1.0&10$^{22}$ & 1&10$^{9}$ & 2&10$^{7}$ & 1&10$^{9}$ \\
 10&$^{-3}$ & 10$^{-5}$ & 10$^{3}$ & 3.5&10$^{21}$ & 6&10$^{8}$ & 3&10$^{8}$ & 1&10$^{9}$ \\
 \hline
\end{tabular}

%% file: table4.tex
\begin{tabular}{r@{}l c c r@{$\times$}l r@{$\times$}l r@{$\times$}l r@{$\times$}l}
 \hline\hline
 \multicolumn{2}{c}{$Z$} & $\chi$ & $P/k$ & \multicolumn{2}{c}{$N_{\rm H_2}$} & \multicolumn{2}{c}{$t_{\rm high}$} & \multicolumn{2}{c}{$t_{\rm min}$} & \multicolumn{2}{c}{$t_{\rm max}$} \\
 \multicolumn{2}{c}{($Z_\odot$)} & (Draine) & (cm$^{-3}$ K) & \multicolumn{2}{c}{(cm$^{-2}$)} & \multicolumn{2}{c}{(yr)} & \multicolumn{2}{c}{(yr)} & \multicolumn{2}{c}{(yr)} \\
 \hline
 10&$^{-1}$ & 10$^{-3}$ & 10$^{1}$ & 1.5&10$^{21}$ & 1&10$^{7}$ & 9&10$^{6}$ & 1&10$^{7}$ \\
 10&$^{-1}$ & 10$^{-3}$ & 10$^{2}$ & 2.8&10$^{21}$ & 6&10$^{7}$ & 8&10$^{6}$ & 1&10$^{8}$ \\
 10&$^{-1}$ & 10$^{-4}$ & 10$^{1}$ & 3.9&10$^{21}$ & 6&10$^{8}$ & 9&10$^{7}$ & 1&10$^{9}$ \\
 \hline
 10&$^{-2}$ & 10$^{-3}$ & 10$^{1}$ & 4.0&10$^{21}$ & 1&10$^{7}$ & 4&10$^{6}$ & 1&10$^{7}$ \\
 10&$^{-2}$ & 10$^{-3}$ & 10$^{2}$ & 5.3&10$^{21}$ & 1&10$^{7}$ & 4&10$^{6}$ & 1&10$^{7}$ \\
 10&$^{-2}$ & 10$^{-3}$ & 10$^{3}$ & 1.3&10$^{21}$ & 2&10$^{7}$ & 6&10$^{6}$ & 1&10$^{8}$ \\
 10&$^{-2}$ & 10$^{-4}$ & 10$^{1}$ & 3.8&10$^{21}$ & 1&10$^{8}$ & 4&10$^{7}$ & 1&10$^{8}$ \\
 10&$^{-2}$ & 10$^{-4}$ & 10$^{2}$ & 5.7&10$^{21}$ & 4&10$^{8}$ & 5&10$^{7}$ & 1&10$^{9}$ \\
 10&$^{-2}$ & 10$^{-5}$ & 10$^{1}$ & 3.4&10$^{21}$ & 1&10$^{9}$ & 4&10$^{8}$ & 1&10$^{9}$ \\
 10&$^{-2}$ & 10$^{-5}$ & 10$^{2}$ & 2.2&10$^{21}$ & 1&10$^{9}$ & 6&10$^{8}$ & 1&10$^{9}$ \\
 \hline
 10&$^{-3}$ & 10$^{-3}$ & 10$^{1}$ & 9.4&10$^{21}$ & 1&10$^{7}$ & 0&10$^{0}$ & 1&10$^{7}$ \\
 10&$^{-3}$ & 10$^{-3}$ & 10$^{2}$ & 8.0&10$^{21}$ & 1&10$^{7}$ & 2&10$^{5}$ & 1&10$^{7}$ \\
 10&$^{-3}$ & 10$^{-3}$ & 10$^{3}$ & 5.1&10$^{21}$ & 9&10$^{6}$ & 3&10$^{6}$ & 1&10$^{7}$ \\
 10&$^{-3}$ & 10$^{-3}$ & 10$^{4}$ & 1.1&10$^{21}$ & 5&10$^{6}$ & 5&10$^{6}$ & 9&10$^{7}$ \\
 10&$^{-3}$ & 10$^{-4}$ & 10$^{1}$ & 1.1&10$^{22}$ & 1&10$^{8}$ & 0&10$^{0}$ & 1&10$^{8}$ \\
 10&$^{-3}$ & 10$^{-4}$ & 10$^{2}$ & 8.4&10$^{21}$ & 1&10$^{8}$ & 2&10$^{6}$ & 1&10$^{8}$ \\
 10&$^{-3}$ & 10$^{-4}$ & 10$^{3}$ & 3.9&10$^{21}$ & 8&10$^{7}$ & 2&10$^{7}$ & 1&10$^{9}$ \\
 10&$^{-3}$ & 10$^{-5}$ & 10$^{1}$ & 1.0&10$^{22}$ & 1&10$^{9}$ & 0&10$^{0}$ & 1&10$^{9}$ \\
 10&$^{-3}$ & 10$^{-5}$ & 10$^{2}$ & 8.3&10$^{21}$ & 1&10$^{9}$ & 8&10$^{6}$ & 1&10$^{9}$ \\
 10&$^{-3}$ & 10$^{-5}$ & 10$^{3}$ & 3.3&10$^{21}$ & 9&10$^{8}$ & 3&10$^{8}$ & 1&10$^{9}$ \\
 \hline
\end{tabular}

%% file: table5.tex
\begin{tabular}{l r@{}l c c r@{$\times$}l r@{$\times$}l r@{$\times$}l r@{$\times$}l}
 \hline\hline
 \multicolumn{1}{c}{$v_{\rm turb}$} & \multicolumn{2}{c}{$Z$} & $\chi$ & $P/k$ & \multicolumn{2}{c}{$N_{\rm H_2}$} & \multicolumn{2}{c}{$t_{\rm high}$} & \multicolumn{2}{c}{$t_{\rm min}$} & \multicolumn{2}{c}{$t_{\rm max}$} \\
 \multicolumn{1}{c}{(km$\,$s$^{-1}$)} & \multicolumn{2}{c}{($Z_\odot$)} & (Draine) & (cm$^{-3}$ K) & \multicolumn{2}{c}{(cm$^{-2}$)} & \multicolumn{2}{c}{(yr)} & \multicolumn{2}{c}{(yr)} & \multicolumn{2}{c}{(yr)} \\
 \hline
 \quad 0.05 & 10&$^{0}$  & 10$^{-5}$ & 10$^{3}$ & 1.8&10$^{21}$ & 2&10$^{8}$ & 2&10$^{8}$ & 1&10$^{9}$ \\
 \quad 0.05 & 10&$^{0}$  & 10$^{-5}$ & 10$^{4}$ & 2.9&10$^{22}$ & 2&10$^{7}$ & 2&10$^{7}$ & 1&10$^{9}$ \\
 \quad 0.05 & 10&$^{-1}$ & 10$^{-5}$ & 10$^{3}$ & 2.1&10$^{21}$ & 1&10$^{9}$ & 9&10$^{8}$ & 1&10$^{9}$ \\
 \quad 0.05 & 10&$^{-1}$ & 10$^{-5}$ & 10$^{4}$ & 7.4&10$^{22}$ & 3&10$^{8}$ & 2&10$^{8}$ & 1&10$^{9}$ \\
 \quad 0.05 & 10&$^{-2}$ & 10$^{-5}$ & 10$^{4}$ & 1.1&10$^{21}$ & 1&10$^{9}$ & 1&10$^{9}$ & 1&10$^{9}$ \\
 \hline
 \quad 0.1  & 10&$^{0}$  & 10$^{-5}$ & 10$^{4}$ & 2.9&10$^{22}$ & 2&10$^{7}$ & 2&10$^{7}$ & 1&10$^{9}$ \\
 \quad 0.1  & 10&$^{-1}$ & 10$^{-5}$ & 10$^{3}$ & 1.5&10$^{21}$ & 1&10$^{9}$ & 1&10$^{9}$ & 1&10$^{9}$ \\
 \hline
\end{tabular}

%% file: 5624.bbl
\begin{thebibliography}{}
\bibitem[\protect\citeauthoryear{Bell et al.}{2006}]{Bell2006} Bell, T.~A., Roueff, E., Viti, S., \& Williams, D.~A. 2006, MNRAS, in press, astro-ph/0607428
\bibitem[\protect\citeauthoryear{Combes}{2000}]{Combes2000} Combes, F. 2000, in Molecular Hydrogen in Space, ed.~F.~Combes, \& G.~Pineau des For\^ets (Cambridge: Cambridge Univ.~Press), 275
\bibitem[\protect\citeauthoryear{Draine}{1978}]{Draine1978} Draine, B.~T. 1978, ApJS, 36, 595
\bibitem[\protect\citeauthoryear{Ellison et al.}{2001}]{Ellison2001} Ellison, S.~L., Yan, L., Hook, I.~M., et al.~2001, A\&A, 379, 393
\bibitem[\protect\citeauthoryear{Giovanelli et al.}{2005}]{Giovanelli2005} Giovanelli, R., Haynes, M.~P., Kent, B.~R., et al.~2005, AJ, 130, 2598
\bibitem[\protect\citeauthoryear{Helfer et al.}{2003}]{Helfer2003} Helfer, T.~T., Thornley, M.~D., Regan, M.~W., et al.~2003, ApJS, 145, 259
\bibitem[\protect\citeauthoryear{Israel}{2005}]{Israel2005} Israel, F.~P. 2005, A\&A, 438, 855
\bibitem[\protect\citeauthoryear{Kalberla et al.}{2000}]{Kalberla2000} Kalberla, P.~M.~W., Kerp, J., \& Haud, U. 2000, in Molecular Hydrogen in Space, ed.~F.~Combes, \& G.~Pineau des For\^ets (Cambridge: Cambridge Univ.~Press), 297
\bibitem[\protect\citeauthoryear{Kennicutt}{1989}]{Kennicutt1989} Kennicutt, R.~C. 1989, ApJ, 344, 685 
\bibitem[\protect\citeauthoryear{Le Teuff et al.}{2000}]{LeTeuff2000} Le Teuff, Y.~H., Millar, T.~J., \& Markwick, A.~J. 2000, A\&AS, 146, 157
\bibitem[\protect\citeauthoryear{Liszt \& Lucas}{1998}]{Liszt1998} Liszt, H.~S., \& Lucas, R. 1998, A\&A, 339, 561
\bibitem[\protect\citeauthoryear{Lloyd}{2003}]{Lloyd2003} Lloyd, C. 2003, in The Calibration Legacy of the ISO Mission, ed.~L.~Metcalfe, A.~Salama, S.~B.~Peschke, \& M.~F.~Kessler (ESA SP-481; Noordwijk: ESA), 399
\bibitem[\protect\citeauthoryear{Lundgren et al.}{2004}]{Lundgren2004} Lundgren, A.~A., Wiklind, T., Olofsson, H., \& Rydbeck, G. 2004, A\&A, 413, 505
\bibitem[\protect\citeauthoryear{Madden}{2002}]{Madden2002} Madden, S.~C. 2002, Ap\&SS, 281, 247
\bibitem[\protect\citeauthoryear{Madden et al.}{1997}]{Madden1997} Madden, S.~C., Poglitsch, A., Geis, N., Stacey, G.~J., \& Townes, C.~H. 1997, ApJ, 483, 200
\bibitem[\protect\citeauthoryear{Noordermeer et al.}{2005}]{Noordermeer2005} Noordermeer, E., van der Hulst, J.~M., Sancisi, R., Swaters, R.~A., \& van Albada, T.~S. 2005, A\&A, 442, 137
\bibitem[\protect\citeauthoryear{Pfenniger et al.}{1994}]{Pfenniger1994} Pfenniger, D., Combes, F., \& Martinet, L. 1994, A\&A, 285, 79
\bibitem[\protect\citeauthoryear{R\"ollig et al.}{2006}]{Rollig2006} R\"ollig, M., Abel, N.~P., Bell, T.~A., et al.~2006, A\&A, submitted
\bibitem[\protect\citeauthoryear{Sargent et al.}{1979}]{Sargent1979} Sargent, W.~L.~W., Young, P.~J., Boksenberg, A., Carswell, R.~F., \& Whelan, J.~A.~J. 1979, ApJ, 230, 49
\bibitem[\protect\citeauthoryear{Smoker et al.}{2000}]{Smoker2000} Smoker, J.~V., Keenan, F.~P., Polatidis, A.~G., et al.~2000, A\&A, 363, 451
\bibitem[\protect\citeauthoryear{Sternberg et al.}{2002}]{Sternberg2002} Sternberg, A., McKee, C.~F., \& Wolfire, M.~G. 2002, ApJS, 143, 419
\bibitem[\protect\citeauthoryear{Walker \& Wardle}{1998}]{Walker1998} Walker, M., \& Wardle, M. 1998, ApJ, 498, L125
\bibitem[\protect\citeauthoryear{Walter et al.}{2005}]{Walter2005} Walter, F., Brinks, E., de Blok, W.~J.~G., Thornley, M.~D., \& Kennicutt, R.~C. 2005, in ASP Conf.~Ser.~331, Extra-Planar Gas, ed.~R.~Braun, 269
\bibitem[\protect\citeauthoryear{Williams et al.}{1995}]{Williams1995} Williams, J.~P., Blitz, L., \& Stark, A.~A. 1995, ApJ, 451, 252
\bibitem[\protect\citeauthoryear{Wilson \& Combes}{1998}]{Wilson1998} Wilson, C.~D., \& Combes, F. 1998, A\&A, 330, 63
\bibitem[\protect\citeauthoryear{Wolfire et al.}{1993}]{Wolfire1993} Wolfire, M.~G., Hollenbach, D., \& Tielens, A.~G.~G.~M. 1993, ApJ, 402, 195
\bibitem[\protect\citeauthoryear{Zwaan \& Prochaska}{2006}]{Zwaan2006} Zwaan, M.~A., \& Prochaska, J.~X. 2006, ApJ, 643, 675
\end{thebibliography}
